\documentclass[pra,preprint,superscriptaddress]{revtex4-1}

\usepackage{amsmath,amssymb}
\usepackage[capitalize]{cleveref}
\usepackage[separate-uncertainty=true,multi-part-units=single]{siunitx}
\usepackage{graphicx}
\usepackage{float}

\providecommand{\config}[3]{\ensuremath{{#1}{#2}^{#3}}}
\providecommand{\term}[3]{\ensuremath{^{#1}\text{#2}_{#3}}}
\providecommand{\ani}[1]{\ensuremath{\text{#1}^{-}}}

\begin{document}
	
	\title{The lifetime of the bound excited level in \ani{Ni}}
	
	\author{M. Kami\'nska}
	\email{magdalena.kaminska@fysik.su.se}
	\affiliation{Department of Physics, Stockholm University, AlbaNova, SE-106 91 Stockholm, Sweden}
		\affiliation{Institute of Physics, Jan Kochanowski University, 25-369 Kielce, Poland}
	\author{V. T. Davis}
	\affiliation{Department of Physics, University of Nevada, Reno, Nevada 89557, United States}
	\author{O. M. Hole}
	\affiliation{Department of Physics, Stockholm University, AlbaNova, SE-106 91 Stockholm, Sweden}
	\author{R. F. Nascimento}
	\affiliation{Department of Physics, Stockholm University, AlbaNova, SE-106 91 Stockholm, Sweden}
	\affiliation{Centro Federal de Educa\c c\~ao Tecnol\'ogica Celso Suckow da Fonseca, Petr\'opolis, 25620-003, RJ, Brazil}​
	\author{KC Chartkunchand}
	\affiliation{Department of Physics, Stockholm University, AlbaNova, SE-106 91 Stockholm, Sweden}
	\affiliation{Department of Physics, University of Nevada, Reno, Nevada 89557, United States}
	\author{M. Blom}
	\author{M. Bj\"{o}rkhage}
	\author{A. K\"{a}llberg}
	\author{P. L\"{o}fgren}
	\author{P. Reinhed}
	\author{S. Ros\'{e}n}
	\author{A. Simonsson}
	\author{R. D. Thomas}
	\author{S. Mannervik}
	\affiliation{Department of Physics, Stockholm University, AlbaNova, SE-106 91 Stockholm, Sweden}
	\author{P. A. Neill}
	\author{J. S. Thompson}
	\affiliation{Department of Physics, University of Nevada, Reno, Nevada 89557, United States}
	\author{H. T. Schmidt}
	\email{schmidt@fysik.su.se}
	\author{H. Cederquist}
	\affiliation{Department of Physics, Stockholm University, AlbaNova, SE-106 91 Stockholm, Sweden}
	\author{D. Hanstorp}
	\affiliation{Department of Physics, University of Gothenburg, SE-412 96 Gothenburg, Sweden}
	
	\begin{abstract}
		The intrinsic lifetime of the upper level in the bound-bound \config{3}{d}{9}\config{4}{s}{2} \term{2}{D}{3/2} $\rightarrow$ \config{3}{d}{9}\config{4}{s}{2} \term{2}{D}{5/2} radiative transition in \ani{Ni} was measured to be \SI{15.1\pm0.4}{\s}. The experiment was performed at cryogenic temperatures in one of the ion-beam storage rings of the DESIREE (Double ElectroStatic Ion Ring Experiment) facility at Stockholm University. The storage lifetime of the \ani{Ni} ion-beam was measured to be close to five minutes at a ring temperature of \SI{13}{\K}.
	\end{abstract}
	
	\date{\today}
	
	\maketitle
	
	Atomic anions are fragile quantum systems with properties that differ significantly from those of atoms or cations. The Coulomb attraction is efficiently shielded making the  1/$r^4$  polarization potential the main interaction between an electron and the neutral atomic core at large distances. The electron correlation is of great importance, and these systems are therefore suitable objects for the testing of theoretical  models that go beyond the independent particle model. Anions also play significant roles in a number of environments.  For instance, physical processes involving negative ions play key roles in astrophysical environments such as stellar and planetary atmospheres, and in the interstellar medium ~\cite{Goudsmit34,Wildt37,Sarre00,Massey40,Wildt44,Warner67,Strittmatter71}. Further, detailed knowledge of energy level schemes and lifetimes of bound and resonant states in atomic anions are  essential in many technological applications, where widely-used radiological dating techniques are the most prominent examples~\cite{Pegg04,Fifield99}. 
		
A direct consequence of the lack of long range attraction is that the number of bound quantum states is strongly limited. Some atoms, such as H and Cu, only form anions with a single bound state~\cite{Andersen04}. In most cases, however, atomic anions have a few bound states below the photodetachment threshold. Then the ground and excited states often have the same parity. The latter states may be higher-lying fine-structure levels or $LS$-terms of either the same~\cite{Pegg04,Andersen99} or a different~\cite{Andersen99,Scheer98} electron configuration as compared with the ground state. Electric dipole transitions are forbidden between states of the same parity and therefore the excited states are long-lived when they have the same parity as the ground state. There are only three known cases of atomic anions with bound excited states of parities \emph{opposite} to those of the ground states, namely \ani{La}~\cite{Walter14}, \ani{Os}~\cite{Bilodeau00,Warring09,Fischer10}, and \ani{Ce}~\cite{Walter07,Walter11}. These systems are of great interest as they may open up long sought-after possibilities of laser cooling and thus sympathetic cooling of negatively charged particles~\cite{Kellerbauer06,Kellerbauer14}.

There are a number of elements where  the attractive potential is so shallow that no bound states can be formed~\cite{Pegg04}. In some of these cases anions instead form metastable states with energies greater than the ground state of the atom. These metastable states mainly decay through non-radiative (electron detachment) processes with lifetimes typically on the order of~microseconds~\cite{Balling92,Andersen93,Reinhed09}.

The topic of this paper is the lifetime of the bound excited state in \ani{Ni}, which decays to the \ani{Ni} ground state through a radiative transition. Theoretical studies of lifetimes of bound excited states in atomic anions are in general sparse due to the large challenges involved in modeling these systems~\cite{Andersen04,Pegg04}. Experimentally, very few studies have been conducted since this requires storage of  negative ions over long times. The first experiments of these types were carried out at the magnetic storage ring CRYRING~\cite{Andersson06,Ellmann04}. However, such measurements are best made in devices that are designed so that stray magnetic fields, which may affect transition rates ~\cite{Pedersen01}, are minimized. The development of the first electrostatic storage ring ELISA ~\cite{Moller97} and the electrostatic  bottle technique by Zajfman~\textit{et al.}~\cite{Zajfman97} are important technological steps forwards in this domain.

Bound excited states in negative ions are sometimes extremely fragile, with binding
energies which may be only on the order of a few \si{\meV}, and therefore require cryogenic storage to avoid detachment by blackbody radiation. The first investigation of negative ions at cryogenic temperatures was demonstrated in the so called CONEtrap device~\cite{Schmidt01,Reinhed09}. This was followed by a cryogenically operated Zajfman trap~\cite{Menk14}, and very recently three cryogenic electrostatic ion storage rings have been put into operation~\cite{Thomas11,vonHahn11,Nakano12}. This latest development opens up the possibility of investigating anions in an entirely new time regime.

As mentioned above, states that are forbidden to decay through single photon electric dipole (E1) transitions within the \emph{LS} coupling approximation are often long-lived and their lifetimes range from \si{\ms} and upwards. If such a state is to undergo a radiative transition to an energetically lower bound state, it must do so via much slower processes such as magnetic and/or higher-order electric multi-pole transitions, or transitions in which more than one photon is emitted.  Examples include: magnetic dipole (M1), electric quadrupole (E2), and/or two-photon transitions (2E1).

Lifetime measurements of long-lived excited \emph{bound} atomic anion states reported up to this date are those for the upper levels in the \config{n}{p}{5} \term{2}{P}{1/2} $\rightarrow$ \config{n}{p}{5} \term{2}{P}{3/2} transitions in~\ani{Te} $(n=5)$~\cite{Andersson06,Ellmann04,Backstrom15}, \ani{Se} $(n=4)$~\cite{Andersson06,Ellmann04,Backstrom15}, and~\ani{S} $(n=3)$~\cite{Backstrom15}. The lifetimes for \ani{Te} and \ani{Se} were first measured in the large magnetic ion storage ring CRYRING~\cite{Andersson06} and found to be in good agreement with multi-configurational Dirac-Fock calculations at the level of the corresponding (limited) experimental accuracy~\cite{Andersson06,Ellmann04}.
	
More recently, the capabilities of the DESIREE facility were demonstrated through more precise measurements of the radiative lifetimes for \ani{Te} and \ani{Se}~\cite{Backstrom15}, and for the long-lived upper level in \ani{S}. In the latter experiment, a \SI{10}{\keV} sulfur anion beam was confined for more than one hour, thus allowing the intrinsic radiative lifetime of its bound excited state to be measured as \SI{503\pm54}{\s}~\cite{Backstrom15}.  This result is in fair agreement with the so far only available theoretical prediction for \ani{S}~\cite{Andersson06}.
	
		\begin{figure}
		\centering
		\vspace{-12pt}
		\includegraphics[width=\columnwidth]{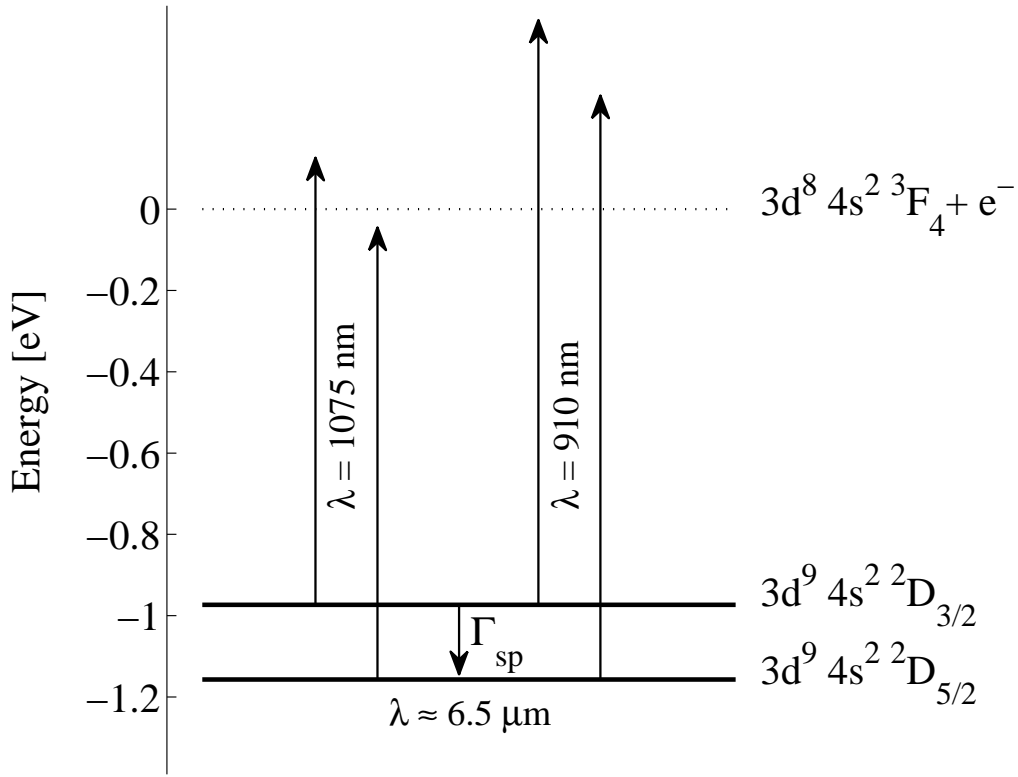}
		\caption{Schematic energy level diagram for the bound states of \ani{Ni}.   The vertical scale indicates energies with respect to the ground state of the neutral Ni atom. In the present work we measure the spontaneous decay rate, $\Gamma_{\text{sp}}$, of the (radiative) \term{2}{D}{3/2} $\rightarrow$ \term{2}{D}{5/2} transition.}	
		\label{EDia}
	\end{figure}
	
To our knowledge, there are no published theoretical predictions of the lifetime of the only bound excited state in \ani{Ni}. The electron affinity of Ni has been measured to be \SI{1157.16(12)}{\meV} and the \config{3}{d}{9}\config{4}{s}{2} \term{2}{D}{3/2} fine-structure excited level lies \SI{184.1(4)}{\meV}  above the \config{3}{d}{9}\config{4}{s}{2} \term{2}{D}{5/2} anionic ground state~\cite{Scheer98}; see~Fig.~\ref{EDia}. 

\begin{figure}
		\centering
		\includegraphics[width=\columnwidth]{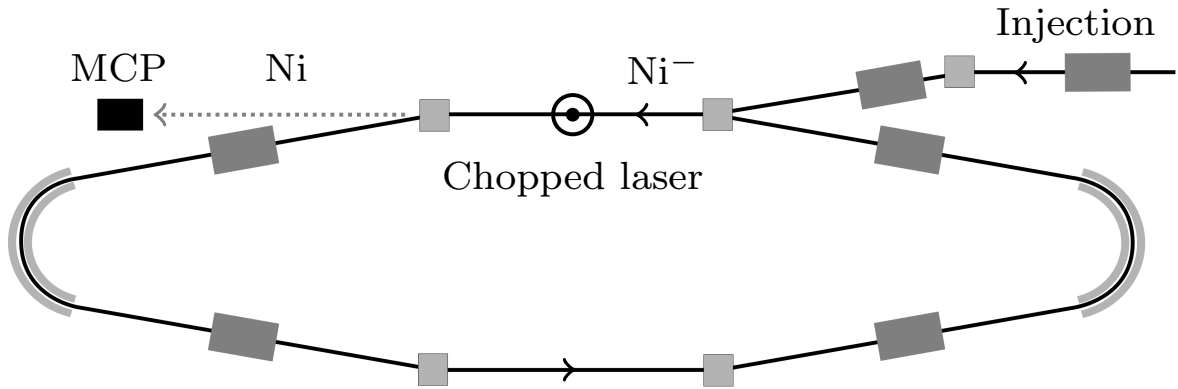}
		\caption{A schematic of the ion storage ring DESIREE. \ani{Ni}~ions (10 keV) are injected and stored in the ring. A~chopped laser beam crosses the ion beam at right angles to yield neutral Ni atoms through electron detachment processes. The neutral atoms are detected by a micro-channel plate (MCP). }
		\label{RING}
	\end{figure}

The present experiment was conducted in the cryogenic storage ring DESIREE. A~full description of this facility, and the procedure for lifetime measurements can be found in references~\cite{Thomas11,Backstrom15,Schmidt13}, so only a~brief description is given here.  Atomic anions of Ni were produced in~a~cesium-sputter ion source (SNICS II)~\cite{SNICS}.  The ions were  accelerated to an energy of~\SI{10}{\keV} to~form an~\SI{\sim10}{\nA} \ani{Ni} ion beam which was injected into the symmetric storage ring, as can be seen in~Fig.~\ref{RING}. The population of the metastable state was probed as~a~function of~storage time by means of photodetachment by~a~chopped laser beam~(Fig.~\ref{RING}). The resulting \SI{10}{\keV}~Ni~atoms were detected by~a~micro-channel plate detector (MCP) operating at cryogenic temperatures~\cite{Thomas11,Schmidt13,Rosen07}. 
	
	A series of mirrors and lenses guided the chopped laser beam through windows in the inner and outer vacuum chambers of DESIREE~\cite{Thomas11} to intersect the stored ion beam at an angle of \ang{90}. The laser intensity at the intersection with the ion beam was of the order of~\si{\mW/\cm^{2}}. The duty cycle of the mechanical shutter, used to control the chopping of the laser beam, was varied between 5\% and 20\% and the effective decay rate was measured as a function of this parameter. This way we established that the influence on the measured decay rate was negligible at the duty cycle applied in the actual measurement~\cite{BackstromDis}. 
	
\begin{figure}
		\centering
		\includegraphics[width=\columnwidth]{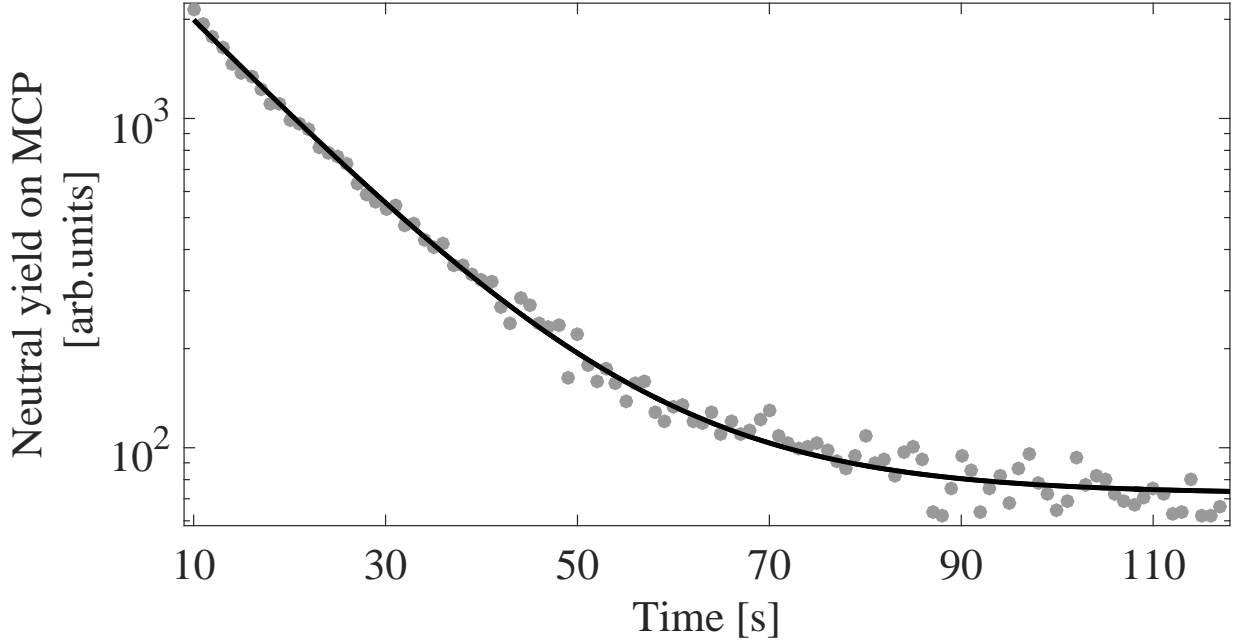}
		\caption{Neutral yield for \ani{Ni} photodetachment at~\SI{1075}{\nm} as a function of time. The full line is a fit to experimental data as described in the text.}
		\label{LPT_MS}
	\end{figure}

	We thus detach the excited state with the chopped laser light for fixed durations at regularly-spaced intervals (the shutter was open for \SI{50}{\ms} at \SI{1}{\s} intervals) after beam injection, and continuously record the neutral atoms with the MCP detector. In~Fig.~\ref{LPT_MS} we show the number of counts during the laser pulse as a function of time after ion injection. An IPG Photonics YLD-5-LP ytterbium fiber laser operating at a wavelength of~\SI{1075}{\nm} was used to photodetach the excited \config{3}{d}{9}\config{4}{s}{2} \term{2}{D}{3/2} ions of \ani{Ni}, while leaving the ground state ions unaffected.

\begin{figure}
		\centering
	\includegraphics[width=\columnwidth]{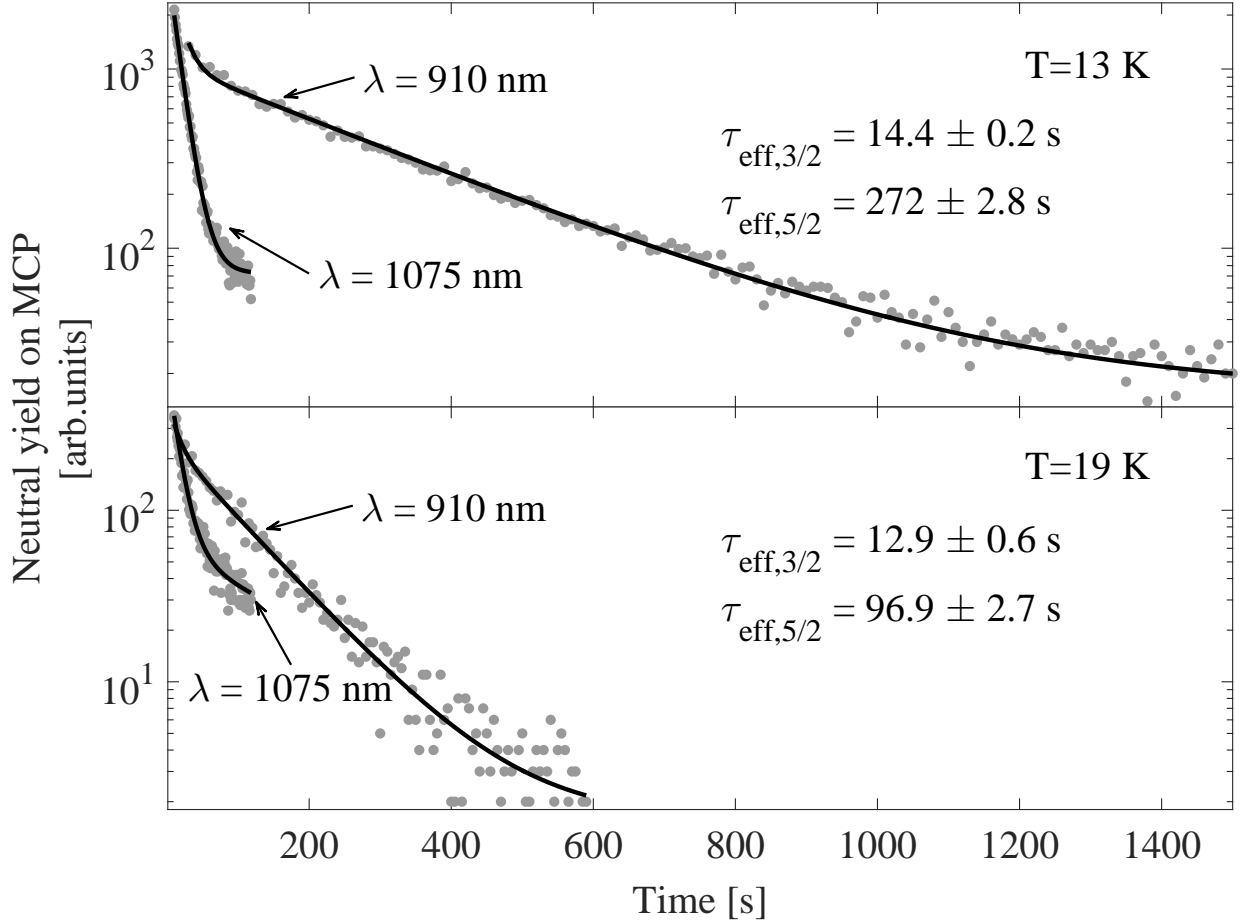}
		\caption{Yields of neutral Ni atoms due to photodetachment of stored \ani{Ni} ions as functions of time. The upper curves in each panel are for laser probing at \SI{910}{\nm} (detachment of the $J=5/2$ and $J=3/2$ levels) while the lower curves are for probing of only the $J=3/2$ level at $\lambda=\SI{1075}{\nm}$. The upper and lower panels show data for ring temperatures of~\SI{13}{\K} and~\SI{19}{\K}, respectively. The full curves are obtained from the fitting procedure described in the text. }
		\label{LTplot}
	\end{figure}

	To measure the effective ion-beam storage lifetime, the above process was repeated using a tunable Ti-sapphire laser operating at a wavelength of \SI{910}{\nm} to photodetach both the anionic ground state and the bound upper fine-structure level. The two data sets in the upper panel of~Fig.~\ref{LTplot} show the integrated number of counts with the laser on for the two different laser wavelengths. For $\lambda=\SI{1075}{\nm}$ only the ions in the metastable state can be photodetached but to a smaller extent even ground state ions will contribute to the signal through collisional detachment in the residual gas. For $\lambda=\SI{910}{\nm}$ both levels can be detached, but the photodetachment cross section is higher for ions in the metastable state than for ground state ions. In both cases the functional form of the time-dependent yield will be a linear combination of $\it{the}$ \emph{same} two exponential functions plus a constant due to the dark count rate of the detector. The two data sets are fitted together in a least-squares fit procedure where the functional forms of the integrated number of counts $N(t)$ and $N'(t)$ at the two wavelengths $\lambda=\SI{1075}{\nm}$ and $\lambda=\SI{910}{\nm}$,  are:

\begin{gather}
	N(t)=A\,\textrm{e}^{(-t/\tau_{\text{eff},3/2})}+B\,\textrm{e}^{(-t/\tau_{\text{eff},5/2})}+C\,n_{\text{inj}}\, \Delta_{\text{ch}}
	\label{eqn:MS} \\
	\intertext{and}
	N'(t) =A'\,\textrm{e}^{(-t/\tau_{\text{eff},3/2})}+B'\,\textrm{e}^{(-t/\tau_{\text{eff},5/2})}+C\,n'_{\text{inj}}\, \Delta'_{\text{ch}}.
	\label{eqn:GS}
\end{gather}

Here, the fit parameters are the four coefficients, $A$, $B$, $A'$, and $B'$, the constant count rate of detector dark counts, $C$, and the two effective lifetimes, $\tau_{\text{eff},3/2}$ and $\tau_{\text{eff},5/2}$. The constants $n_{\text{inj}}$, $\Delta_{\text{ch}}$, $n'_{\text{inj}}$, and $\Delta'_{\text{ch}}$ are the numbers of injections and time-increments per channel for the data sets at~\SI{1075}{\nm} and \SI{910}{\nm}, respectively. By this procedure the two  effective lifetimes are extracted. These are sensitive to both data sets, but the shorter one, $\tau_{\text{eff},3/2}$, is most sensitive to the data recorded at~\SI{1075}{\nm}, while the longer one, $\tau_{\text{eff},5/2}$, is mainly determined by the data recorded at \SI{910}{\nm} (see~Fig.~\ref{LTplot}). As the background is described through the constant dark count rate, $C$, it is not necessary to extend the measurement at~\SI{1075}{\nm} to long times to determine the background. In this way, time can instead be spent more efficiently by making more injections.

\begin{figure}
		\centering
		\includegraphics[width=\columnwidth]{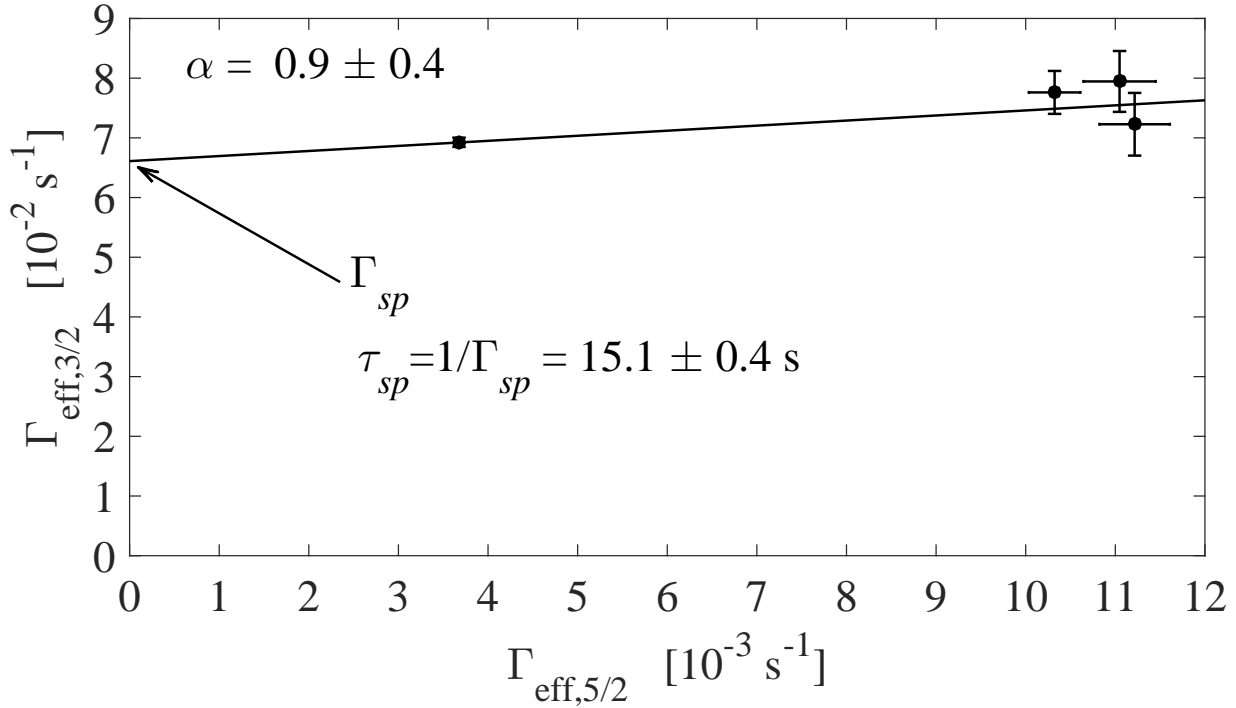}
		\caption{Stern-Vollmer plot of the effective decay rate of the \term{2}{D}{3/2} level plotted as a function of the effective decay rate of the \term{2}{D}{5/2} level (the ground state). The line is a fit to the data and its intercept with the vertical axis ($\Gamma_{5/2}=0$) gives the spontaneous decay rate $\Gamma_{\text{sp}}$ as collisions with the residual gas dominate the beam losses.}
		\label{SVplots}
	\end{figure}

The effective lifetimes extracted from the data sets recorded at the base temperature of \SI{13}{\K} are \mbox{$\tau_{\text{eff},3/2}=\SI{14.4\pm0.2}{\s}$} and $\tau_{\text{eff},5/2}=\SI{272.0\pm2.8}{\s}$. As we will argue below, the effective storage lifetime ($\tau_{\text{eff},5/2}$) is to a very good approximation only limited by residual gas collisions. We can then assume that it is inversely proportional to the residual-gas density. The effective decay rate of the short component $\Gamma_{\text{eff},3/2}=1/\tau_{\text{eff},3/2}$ is under these conditions a sum of~the~spontaneous decay rate to be determined and a term proportional to the residual-gas density (which is about 10$^4$~H$_2$~molecules per cm$^3$~\cite{Schmidt13}). By measuring at different residual-gas densities we obtain pairs of~$\tau_{\text{eff},3/2}$~-~against $\tau_{\text{eff},5/2}$ - values and deduce the decay rate in the absence of residual-gas collisions by extrapolating to $\Gamma_{\text{eff},5/2}$=0, as shown in~Fig.~\ref{SVplots}. To control the residual-gas density we raised the temperature of the cryogenic vacuum chamber slightly (to~\SI{19}{\K}) and repeated the measurements at~both wavelengths. An~example of such measurements at \SI{19}{\K} yielded the data in the lower panel of~Fig.~\ref{LTplot}. 

We have used several sets of measurements at the elevated temperature of \SI{19}{\K}. The reason for this is~that the residual gas density at this temperature may depend somewhat on the rate of temperature increase (from~\SI{13}{\K}) and earlier heating/baking cycles. The resulting effective decay rates at \SI{13}{\K} and \SI{19}{\K} are plotted in the Stern-Vollmer plot in~Fig.~\ref{SVplots}, where the spread of~the~\SI{19}{\K} data points reflects the uncertainties discussed above. The line is a least-squares fit to the data and the spontaneous decay rate corresponds to an intrinsic  lifetime of $\tau_\text{sp}$=1/$\Gamma_{\text{sp}}$=\SI{15.1\pm0.4}{\s}. The slope of the line is~$\alpha=0.9\pm0.4$. This is consistent with $\alpha=1$, which would be the result if the cross sections for collisional detachment were identical for the $J=3/2$ and $J=5/2$ levels. It is important to stress that our result is obtained without making any assumption of the value of $\alpha$ and instead we use the linear fit in~Fig.~\ref{SVplots}. 

Our result $\tau_\text{sp}$=\SI{15.1\pm0.4}{\s} is obtained by considering that losses other than those due to collisions in the residual gas are unimportant. In connection with the measurement of the lifetime of the excited bound state in \ani{S}~\cite{Backstrom15}, it was concluded  that the beam loss rate in the absence of residual gas must have been smaller than 10$^{-3}$~s$^{-1}$~\cite{Backstrom15}. As the same ion beam energy, \SI{10}{\keV}, was used in that experiment~\cite{Backstrom15} as here, the rate for \ani{Ni} beam losses without any residual gas would be less than 10$^{-3}$~s$^{-1}$. Furthermore, such losses should be equally important for ions in the ground and metastable states. Therefore they would shift the four data points in~Fig.~\ref{SVplots} by equal amounts along the vertical and horizontal axes. As the slope of the straight line fit is close to unity, the data points would basically just shift along this line. This could only lead to corrections much smaller than the error we give in our final result.

No theoretical results for the lifetime measured here are currently available but calculations are underway by Beerwerth and Fritsche~\cite{Fritsche15}. We hope that these new precision measurements of lifetimes of bound excited states in negative ions will also initiate further theoretical investigations. It is well known that such calculations are challenging. As an example, O'Malley and Beck~\cite{O'Malley03} have reported lifetimes of~\SI{162}{\s} and \SI{27.3}{\hour} for the $J=3/2$ and $J=5/2$ levels of~the~\term{2}{D}{}~term in \ani{Si} using relativistic configuration interaction calculations.  Froese Fischer, on the other hand  obtained \SI{14.5}{\hour} and \SI{12.2}{\hour}~\cite{Andersson06} for the same levels using Multi Configuration Dirac-Hartree-Fock calculations~\cite{Tachiev99}. Clearly, experimental data are needed in order to test the models used to describe decay processes in negative ions.

	In conclusion, we report the lifetime of the excited fine-structure level in \ani{Ni} to be \SI{15.1\pm0.4}{\s}.  The relatively small error is due to the fact that the ion-beam storage lifetime is almost a factor of twenty longer than the excited state lifetime.  This is in contrast to the much longer \ani{S} lifetime that we recently reported~\cite{Backstrom15}, in which the beam storage lifetime was only about twice as long as the spontaneous radiative decay lifetime.  
	
In order to improve the precision in the lifetime measurements, we plan to make separate measurements of~the~absolute detachment cross-sections for ground and excited state \ani{Ni} ions colliding with H$_2$ in a cell on the ion-injection beam line of DESIREE~\cite{Thomas11,Schmidt13}. Furthermore, we plan to perform measurements of lifetimes of bound excited states in atomic anions starting with~\ani{Pt}~\cite{Bilodeau99} and \ani{In}~\cite{Walter10}.	
	
	This work was supported by the Swedish Research Council (Contracts No. 821-2013-1642, No. 621-2012- 3662, No. 621-2014-4501, and No. 621-2013-4084) and by the Knut and Alice Wallenberg Foundation. We acknowledge support from the COST action CM1204 XUV/X-ray light and fast ions for ultrafast chemistry (XLIC). MK acknowledges financial support from the Mobility Plus (project 1302/MOB/IV/2015/0) founded by the Polish Ministry of Science and Higher Education.

	\bibliography{NiBib}
	
\end{document}